\documentclass{article}

\usepackage[english]{babel}

\usepackage[letterpaper,top=2cm,bottom=2cm,left=3cm,right=3cm,marginparwidth=1.75cm]{geometry}

\usepackage{amsmath}
\usepackage[ruled,linesnumbered]{algorithm2e}
\usepackage{graphicx}
\usepackage[colorlinks=true, allcolors=blue]{hyperref}
\usepackage[noend]{algpseudocode}
\usepackage{multirow}

\title{A Multi-parameter Updating Fourier Online Gradient Descent Algorithm for Large-scale Nonlinear Classification}
\author{Yingying Chen}

\begin{document}
\maketitle

\begin{abstract}
Large scale nonlinear classification is a challenging task in the field of support vector machine. Online random Fourier feature map algorithms are very important methods for dealing with large scale nonlinear classification problems. The main shortcomings of these methods are as follows: (1) Since only the hyperplane vector is updated during learning while the random directions are fixed, there is no guarantee that these online methods can adapt to the change of data distribution when the data is coming one by one. (2) The dimension of the random direction is often higher for obtaining better classification accuracy, which results in longer test time. In order to overcome these shortcomings, a multi-parameter updating Fourier online gradient descent algorithm (MPU-FOGD) is proposed for large-scale nonlinear classification problems based on a novel random feature map. In the proposed method, the suggested random feature map has lower dimension while the multi-parameter updating strategy can guarantee the learning model can better adapt to the change of data distribution when the data is coming one by one. Theoretically, it is proved that compared with the existing random Fourier feature maps, the proposed random feature map can give a tighter error bound. Empirical studies on several benchmark data sets demonstrate that compared with the state-of-the-art online random Fourier feature map methods, the proposed MPU-FOGD can obtain better test accuracy. 
\end{abstract}

\section{Introduction}

Support vector machines (SVM) are powerful tools for data classification in machine learning and have been widely applied into the fields of pattern recognition \cite{zhao2016person,chang2016semantic,hare38s,dhamecha2019between,ye2019hybrid,nanni2017handcrafted}, image processing \cite{yao2018extracting,guo2018effective,zuo2017distance}, computer vision \cite{kang2019grayscale,zhang2019visual}, data mining \cite{wen2018efficient,xu2018modeling}, and so on. Generally speaking, the computational complexity of SVM is  $O(n^3)$, which leads to that training SVM becomes a challenging task for large scale classification problems.

At present, the existing methods for large scale classification problems can be categorized into two classes: offline learning and online learning. In the case of offline learning, on the one hand, the researchers have proposed Lagrangian support vector machine algorithm \cite{mangasarian2001lagrangian}, stochastic gradient descent algorithms \cite{zhang2004solving,Shalev2007Pegasos}, modified finite Newton algorithm \cite{keerthi2005modified}, cutting-plane method \cite{joachims2006training}, Bundle method \cite{le2008bundle,teo2010bundle}, coordinate descent method \cite{chang2008coordinate}, and dual coordinate descent method \cite{hsieh2008dual} to deal with large scale linear classification problems. On the other hand, the researchers have presented SMO algorithm \cite{keerthi2001improvements}, low-rank kernel representation method \cite{smola2000sparse,fine2001efficient}, reduced support vector machine \cite{lee2001rsvm}, core vector machine \cite{tsang2005core}, and localized support vector machine \cite{zhang2006svm,cheng2009efficient,chang2010tree} to deal with large scale nonlinear classification problems. In machine learning, research on algorithms for large scale linear classification problems has been mature. Inspired by this fact, a finite-dimensional mapping based offline algorithm has been proposed to deal with large scale nonlinear classification problems \cite{rahimi2008random}. Specifically, for the positive definite shift-invariant kernel, it first makes use of random Fourier features to map the input data to a randomized low-dimensional feature space, and then uses linear algorithms to deal with large scale nonlinear classification problems. 

In the previous studies, the idea of converting a batch optimization problem into an online task was suggested in \cite{li2000relaxed}. Inspired by \cite{li2000relaxed} and \cite{rosenblatt1958perceptron}, an online learning framework \cite{crammer2006online} was suggested based on the passive-aggressive strategy to deal with linear classification, linear regression, and uniclass prediction problems. The main difference between \cite{rosenblatt1958perceptron} and \cite{crammer2006online} is that except for classification and regression, \cite{crammer2006online} can deal with uniclass prediction problem. For large-scale nonlinear classification problems, \cite{wang2013large} extended \cite{rahimi2008random} to online learning and proposed Fourier online gradient descent algorithm (FOGD), which often needs large number of random features $D$. \cite{lin2014sample} answered the question "How many random Fourier features are needed to obtain the better results in the online kernel learning setting?". A local online learning algorithm was proposed \cite{zhou2016one} based on the observation that although the data sets are not globally linear separable, they may still be locally linear separable. Using the reparameterized random feature, a large-scale online kernel learning algorithm was proposed \cite{nguyen2017large}. \cite{jorge2018passive} suggested an online learning algorithm via combining the passive-aggressive strategy and max-out function (PAMO). Using the Nyström approximation technology, an online gradient descent (NOGD) algorithm was designed \cite{lu2016large}. Unfortunately, it is not easy to approximate the whole kernel matrix using a submatrix. The smaller submatrix will degrade test accuracy while the larger submatrix will lead to higher computational complexity. For the random Fourier feature mapping based online learning, generally speaking, a larger $D$ leads to a more precise approximation. However, a larger $D$ leads to a higher computational complexity. To reduce the impact number of random features, recently, a dual space gradient descent algorithm (DualSGD) was proposed \cite{le2016dual}.

The main shortcomings of random Fourier feature mapping based online learning methods are as follows: (1) Since only the hyperplane vector is updated during learning while the random directions are fixed, there is no guarantee that these online methods can adapt to the change of data distribution when the data is coming one by one. (2) The dimension of the random direction is often higher for obtaining better classification accuracy, which results in longer test time. In order to overcome these shortcomings, in this study, a multi-parameter updating Fourier online gradient descent algorithm (MPU-FOGD) is proposed for large-scale nonlinear classification problems based on a novel random feature map. In the proposed method, the suggested random feature map has lower dimension while the multi-parameter updating strategy can guarantee the learning model can better adapt to the change of data distribution when the data is coming one by one. Theoretically, we prove that the proposed random feature map can give a tighter error bound. Empirical studies on several benchmark data sets show that compared with the state-of-the-art online random Fourier feature map methods, the proposed MPU-FOGD can obtain better test accuracy.

The rest of this paper is organized as follows. Section 2 describes the proposed feature map in details and presents theoretical guarantees. Section 3 gives experiment results and analysis. Section 4 encloses our paper with future work.

\section{MPU-FOGD}

\subsection{Randomized Fourier Feature Map Algorithm}
Let $m_t=\{x_t,lb_t\}, t=1,2,\cdots,n$ be the training examples, where $n$ is the number of examples, $x_t\in R^d$ is the input data, $lb_t\in \{-1,+1\}$ and $lb_t\in \{1,2,\cdots,m\}$ are the corresponding label for binary classification and multi-class classification, respectively. Let $H$ be a Reproducing Kernel Hilbert Space (RKHS) endowed with a kernel function $k(\cdot, \cdot):R^d\times R^d\to R$.

For binary classification problems, the standard model of SVM is as follows: 
\begin{equation}
\min \limits_{w,b} \frac{\lambda}2||w||_H^2+\sum _{t=1}^nl(f(x_t);lb_t),
\end{equation}
where $\lambda >0$, $f(x_t)=w^T \varphi (x_t)+d$, $l(lb_t; f(x_t))=max(0,1-lb_tf(x_t))$. 

The computational complexity of this model is usually $O(n^3)$. For large scale nonlinear classification problems, it undoubtedly faces  computational difficulty. Considering that research on SVM algorithms for large scale linear classification problems has been mature in machine learning, a randomized Fourier feature map $(\varphi (x))^T=(cos(u_1^Tx),cos(u_2^Tx),
\cdots, cos(u_D^Tx)),sin(u_1^Tx),sin(u_2^Tx),\cdots,sin(u_D^Tx))$
is suggested in \cite{rahimi2008random} to breakthrough this bottleneck, where $D=O(d\epsilon ^{-2}log(\frac1{\epsilon ^2}))$, $\epsilon$ is the accuracy of approximation, the random vectors $u_i$ for $\forall i \in \{1,2,\cdots,D\}$ are drawn from 
\begin{equation*}
p(u)=(\frac 1{2\pi})^d \int e^{-iu^T(\Delta x)}d(\Delta x).
\end{equation*}
Obviously, using the trick in \cite{rahimi2008random}, the original data are mapped into a finite low-dimensional feature space and linear learning algorithms can be used to deal with large scale nonlinear classification problems. This is a very important contribution for large scale nonlinear classification problems in the field of machine learning.

\subsection{Proposed Random Fourier Feature Map}
Let $z(x,u)^T=(cos(u^Tx),sin(u^Tx))$, where $z(x)^Tz(y)=\frac {1}{D}\sum _{j=1}^{D}z(x,u_j)z(y,u_j)$.
According to Bochner's theorem \cite{guo2018effective}, we know that $z(x)^T z(y)$ is an unbiased estimate of the positive definite shift-invariant kernel $k(x,y)$ \cite{rahimi2008random}. From \textbf{Claim 1} in \cite{rahimi2008random}, we know that the discrepancy between $k(x,y)$ and $z(x)^T z(y)$ can be narrowed down by varying $D$. Generally speaking, a larger $D$ will lead to a more precise approximation, but it will generate a higher computational complexity. Let $z_{old}(x)^Tz_{old}(y)=\frac {1}{D}\sum _{j=1}^{D}z_{old}(x,u_j,b_j)z_{old}(y,u_j,b_j)$, where $z_{old}(x,u,b)=\sqrt 2cos(u^Tx+b)$, $b$ is drawn uniformly from $[0,2\pi]$. From \cite{rahimi2008random}, we know that $E(z_{old}(x)^Tz_{old}(y))=k(x,y)$, seeing Appendix A for details. Let $z_{new}(x)^Tz_{new}(y)=\frac {1}{D}\sum _{j=1}^{D}z_{new}(x,u_j,b_j)z_{new}(y,u_j,b_j)$, where $z_{new}(x,u,b)=\sqrt \frac 2Dcos(u^Tx+b)$. In order to show that $z_{new} (x)^T z_{new} (y)$  is also an unbiased estimate of $k(x,y)$, we give the following Lemmas.

Based on the above analysis,$(\varphi (x))^T=(\frac {\sqrt 2}{D}cos(u_1^T x+b_1),\frac {\sqrt 2}{D}cos(u_2^T x+b_2),\cdots,\frac {\sqrt 2}{D}cos(u_D^T x+b_D))$ is adapted in this paper. Theoretically, this map will be able to obtain the better accuracy.

\textbf{\subsection{Proposed Algorithm}}

In the existing work, what non-stationary streaming learning environments \cite{gama2004learning,frias2014online,pesaranghader2016fast} emphasize more is whether the change of posterior probability causes the change of its classification decision boundary. Learning from streaming data with concept drift, they usually have a detector to detect whether the distribution of data has changed. No matter what kind of detector they use, if the error rate changes significantly, then concept drift occurs, otherwise the current single classifier is used to predict examples and update model. On the other hand, Fourier feature mapping is obviously data independent. In the previous study, only the hyperplane vector $w$ is updated during learning while the random directions $u$ and $b$ are fixed. Obviously, there is no guarantee that the random Fourier-feature-mapping-based large scale online learning approaches could adapt to the change of data distribution. So, in this study, inspired by streaming learning, we iteratively update random directions $u$ and $b$ which can change the data distribution in the feature space and make the data better fit in with the current space.

MPU-FOGD includes two algorithms, i.e. MPU-FOGDU and MPU-FOGDUB.
In comparison with the algorithm of MPU-FOGDUB, MPU-FOGDU only
updates Fourier component $u$.
Following the framework of large scale online kernel learning \cite{wang2013large}, when the loss function $l(lb_t, w_t^T\varphi (x_t))>0$ for the $t$-th incoming sample, the update formulas of $u_{t+1}$, $b_{t+1}$ and $w_{t+1}$ in the proposed MPU-FOGD are as follows:
\begin{align}
&w_{t+1}=w_t - \eta _1 \nabla _{w_t}l(lb_t, w_t^T\varphi (x_t)),\\
&u_{t+1}=u_t - \eta _2 \nabla _{u_t}l(lb_t, w_t^T\varphi (x_t)),\\
&b_{t+1}=b_t - \eta _3 \nabla _{b_t}l(lb_t, w_t^T\varphi (x_t)),
\end{align}
where
\begin{equation}
\begin{aligned}
&\nabla _{w_t}l(lb_t, w_t^T\varphi (x_t)) = -lb_t\varphi (x_t),\\
\end{aligned}
\end{equation}
\begin{equation}
\begin{aligned}
\nabla _{u_t}l(lb_t, w_t^T\varphi (x_t)) 
%\end{aligned}
%\end{equation}
%\begin{equation}
%\begin{aligned}
=&\frac {\sqrt 2}{D} lb_tw_t^T*(sin(u_{t,1}^Tx_t+b_{t,1}),sin(u_{t,2}^Tx_t+b_{t,2}),\\
&\cdots,sin(u_{t,D}^Tx_t+b_{t,D}))
*\underbrace{(x_t,x_t,\cdots,x_t)}_{D},
\end{aligned}
\end{equation}
\begin{equation}
\begin{aligned}
&\nabla _{b_t}l(lb_t, w_t^T\varphi (x_t)) 
= \frac {\sqrt 2}{D} lb_tw_t^T*(sin(u_{t,1}^Tx_t+b_{t,1}),sin(u_{t,2}^Tx_t+b_{t,2}),\cdots,sin(u_{t,D}^Tx_t+b_{t,D})),
\end{aligned}
\end{equation}             
where * denotes dot product. 

Based on the analysis mentioned above, the detailed steps of the proposed MPU-FOGDUB algorithm for binary classification are given in Algorithm 1.

\begin{algorithm}
	\caption{the MPU-FOGDUB algorithm}\label{algorithm}
	\KwData{D , $\eta$, $\sigma$ }
	\KwOut{model: $w_{t+1}\in R^D$, $u_{t+1}\in R^{d\times D},b_{t+1}\in R^D$ }
	initial $w_0=0$\;
	initial $u_{0,i}$ for $\forall i \in \{1,2,\cdots,D\}$ are drawn from $Dp(u)$\;
	initial $b_{0,i}$ for $\forall i \in \{1,2,\cdots,D\}$ are drawn from $[0, 2\pi]$\;
	%	$\textbf{b}$ is drawn from $[0,2\pi]$\;
	\ForAll{t of $[T]$}{
		receive instance: $x_t\in R^d$\; 
		compute instance representation:\\ $(\varphi (x_t))^T=(\frac {\sqrt 2}{D}cos(u_{t,1}^T x_t+b_{t,1}),\frac {\sqrt 2}{D}cos(u_{t,2}^T x_t+b_{t,2}),\cdots,\frac {\sqrt 2}{D}cos(u_{t,D}^T x_t+b_{t,D}))$\;
		predict $\hat{lb}_t=sign(w_t^T\varphi (x_t))$\;
		receive correct label: $lb_t\in \{-1,1\}$\;
		compute hinge loss:\\
		$l(lb_t, w_t^T\varphi (x_t))=max\{0,1-lb_t(w_t^{T}\varphi (x_t))\}$\;
		\If{ $l(lb_t, w_t^T\varphi (x_t))>0$}{
			$w_{t+1}=w_t-\eta _1\nabla _{w_t}l(lb_t, w_t^T\varphi (x_t))$\;
			$u_{t+1}=u_t-\eta _2\nabla _{u_t}l(lb_t, w_{t}^T\varphi (x_t))$\;  
			$b_{t+1}=b_t-\eta _3\nabla _{b_t}l(lb_t, w_{t}^T\varphi (x_t))$;}
	}
\end{algorithm}

In the multi-class problem setting, the algorithm predicts a sequence of scores for the $m$ classes:
\begin{equation}
\begin{aligned}
(f_{t,1}(x_t), f_{t,2}(x_t),\cdots,f_{t,m}(x_t)).
\end{aligned}
\end{equation}
The hinge loss in \cite{lu2016large} function as follows:
\begin{equation}
\begin{aligned}
l(\gamma _t)=max(0,1-\gamma _t),
\end{aligned}
\end{equation}
where $\gamma _t=f_{t,lb_t}(x_t)-f_{t,s_t}(x_t)$, $s_t=argmax _{r\in Y,r\neq lb_t}f_{t,r}(x_t)$.
%The prediction label is set to the category with the highest prediction score:
%\begin{equation}
%\begin{aligned}
%\hat {lb}_t=argmax _{r\in Y}f_{t,r}(x_t).
%\end{aligned}
%\end{equation}

\textbf{\section{Experiments}}

In this section, we empirically validate performance of the proposed algorithm over various benchmark data sets which can be freely downloaded from the LIBSVM$\footnote{https://www.csie.ntu.edu.tw/~cjlin/libsvmtools/datasets/}$ and UCI$\footnote{https://archive.ics.uci.edu/ml/datasets.php}$ website. Table 1 shows the details of 12 publicly available data sets. Mnist600k is obtained by randomly extracting 600,000 samples from the mnist8m$\footnotemark[1]$ data set after dimension reduction and normalization. All algorithms are implemented in Python 3.7.3, on a windows machine with AMD Ryzen 5 2600 Six-Core Processor@ 3.4GHZ. The codes of the proposed algorithm are written by using the online learning algorithm library \cite{hoi2014libol}.
%Intel® Xeon® CPU E5-2680 v4@ 2.40GHZ

%\begin{table}
%\centering
%	\begin{tabular}{l|rrrr}
%		\hline
%		Dataset & Instances & Features & Classes \\
%		\hline
%		KDDCUP08 & 102,294 & 117 & 2 \\
%		ijcnn1 & 141,691 & 22 & 2 \\
%		codrna & 271,617 & 8 & 2 \\
%		w7a & 24,692 & 300 & 2 \\		
%		skin & 245,057 & 4 & 2 \\
%		covtype & 581,012 & 54 & 2 \\
%		poker & 1,000,000 & 10 & 10 \\
%		mnist600k & 600,000 & 50 & 10 \\
%		forest & 581,012 & 54 & 7 \\
%		acoustic & 98,528 & 50 & 3 \\
%		aloi & 108,000 & 128 & 1000 \\
%		combined & 98,528 & 100 & 3 \\
%		\hline
%	\end{tabular}
%\caption{\label{tab:data}{Data sets used in our experiments.}
%\end{table}
%%\end{center}

\begin{table}[htbp]
	\centering
	\begin{tabular}{l|rrrrr}
		\hline
		Dataset & Instances & Features & Classes \\
		\hline
		KDDCUP08 & 102,294 & 117 & 2 \\
		ijcnn1 & 141,691 & 22 & 2 \\
		codrna & 271,617 & 8 & 2 \\
		w7a & 24,692 & 300 & 2 \\		
		skin & 245,057 & 4 & 2 \\
		covtype & 581,012 & 54 & 2 \\
		poker & 1,000,000 & 10 & 10 \\
		mnist600k & 600,000 & 50 & 10 \\
		forest & 581,012 & 54 & 7 \\
		acoustic & 98,528 & 50 & 3 \\
		aloi & 108,000 & 128 & 1000 \\
		combined & 98,528 & 100 & 3 \\
		\hline
	\end{tabular}
	\caption{\label{tab:widgets}An example table.}
\end{table}

\begin{table}
	\centering
%	\caption{\textbf{Table 2:} Comparison of training time, test time and test accuracy on the binary classification tasks}
	% Table generated by Excel2LaTeX from sheet '统一参数全部算法'
		\begin{tabular}{l|cccccc}
	\hline
% Table generated by Excel2LaTeX from sheet 'Sheet1'
		\multirow{2}[0]{*}{Algorithm} & \multicolumn{3}{c}{skin} & \multicolumn{3}{c}{kddcup08} \\
& Train(s) & Test(s) & TestAcc(\%) & Train(s) & Test(s) & TestAcc(\%) \\
		\hline
		LR &       2.98 &       0.44 &    90.233  &      1.22  &      0.18  &   100.000  \\
		
		OGD &       2.51 &       0.38 &    90.068  &      0.80  &      0.16  &   100.000  \\
		
		PA &       5.15 &       0.38 &    81.630  &      1.96  &      0.16  &   100.000  \\
%		\hline
		FOGD &      15.44 &       3.52 &    99.942  &      2.39  &      0.49  &   100.000  \\
		
		NOGD &      75.12 &       7.06 &    99.047  &     28.29  &      2.79  &    99.452  \\
		
		RRF &       9.98 &       2.38 &    99.949  &      2.40  &      0.48  &   100.000  \\
		
		PAMO &      99.58 &      50.78 &    99.829  &      5.05  &      1.29  &   100.000  \\
		
		AVM &      10.12 &       1.32 &    99.616  &    191.03  &      1.14  &    99.717  \\
		
		MPU-FOGDU &       5.83 &       1.03 &    99.951  &      2.59  &      0.47  &   100.000  \\
		
		MPU-FOGDUB &       6.26 &       1.11 &    99.954  &      2.69  &      0.47  &   100.000  \\
		\hline
		\multirow{2}[0]{*}{Algorithm} & \multicolumn{3}{c}{ijcnn1} & \multicolumn{3}{c}{w7a} \\
& Train(s) & Test(s) & TestAcc(\%) & Train(s) & Test(s) & TestAcc(\%) \\
%		\multicolumn{1}{|c|}{\multirow{2}[4]{*}{Algorithm}} & \multicolumn{3}{c|}{ijcnn1} & \multicolumn{3}{c|}{w7a} \\
%		\cline{2-7}    & \multicolumn{1}{c|}{Train(s)} & \multicolumn{1}{c|}{Test(s)} & \multicolumn{1}{c|}{TestAcc} & \multicolumn{1}{c|}{Train(s)} & \multicolumn{1}{c|}{Test(s)} & \multicolumn{1}{c|}{TestAcc} \\
		\hline
		LR &      1.72  &      0.26  &    67.672  &      0.32  &      0.05  &    95.556  \\
		
		OGD &      1.72  &      0.22  &    71.670  &      0.26  &      0.05  &    96.457  \\
		
		PA &      3.34  &      0.22  &    70.022  &      0.61  &      0.05  &    95.681  \\
%		\hline
		FOGD &     11.16  &      2.59  &    98.014  &      8.34  &      2.04  &    98.477  \\
		
		NOGD &     15.46  &      1.78  &    90.572  &     52.33  &      3.92  &    97.062  \\
		
		RRF &     18.83  &      3.63  &    97.780  &    208.44  &     37.91  &    98.494  \\
		
		PAMO &    191.55  &     43.60  &    98.687  &     28.62  &      6.25  &    98.664  \\
		
		AVM &    430.71  &      1.43  &    94.062  &     35.48  &      4.90  &    98.360  \\
		
		MPU-FOGDU &     21.29  &      3.01  &    98.885  &    25.23  &      0.55  &    98.388  \\
		
		MPU-FOGDUB &     23.35  &      3.00  &    98.885  &     24.77  &      0.54  &    98.396  \\
		\hline
		\multirow{2}[0]{*}{Algorithm} & \multicolumn{3}{c}{codrna} & \multicolumn{3}{c}{covtype} \\
& Train(s) & Test(s) & TestAcc(\%) & Train(s) & Test(s) & TestAcc(\%) \\
%		\multicolumn{1}{|c|}{\multirow{2}[4]{*}{Algorithm}} & \multicolumn{3}{c|}{codrna} & \multicolumn{3}{c|}{covtype} \\
%		\cline{2-7}    & \multicolumn{1}{c|}{Train(s)} & \multicolumn{1}{c|}{Test(s)} & \multicolumn{1}{c|}{TestAcc} & \multicolumn{1}{c|}{Train(s)} & \multicolumn{1}{c|}{Test(s)} & \multicolumn{1}{c|}{TestAcc} \\
		\hline
		LR &      3.27  &      0.49  &    91.790  &      7.13  &      1.07  &    75.317  \\
		
		OGD &      2.71  &      0.43  &    91.843  &      6.52  &      0.95  &    76.072  \\
		
		PA &      6.48  &      0.43  &    91.791  &     13.88  &      0.94  &    75.195  \\
%		\hline
		FOGD &      5.46  &      1.07  &    96.172  &    180.97  &     42.36  &    85.316  \\
		
		NOGD &     92.45  &      8.17  &    92.468  &    336.65  &     29.78  &    71.721  \\
		
		RRF &      8.78  &      1.21  &    96.030  &   2936.24  &    214.44  &    85.045  \\
		
		PAMO &    107.18  &     20.93  &    96.258  &    101.66  &     10.62  &    81.204  \\
		
		AVM &     49.50  &      2.83  &    95.107  &   1418.75  &      7.51  &    71.149  \\
		
		MPU-FOGDU &     11.42  &      1.76  &    96.588  &   1353.59  &     20.21  &    90.880  \\
		
		MPU-FOGDUB &     37.39  &      5.59  &    96.701  &   1419.69  &     19.91  &    90.927  \\
		\hline
	\end{tabular}%
\caption{\label{tab:binary}Comparison of training time, test time and test accuracy on the binary classification tasks.}
\end{table} 
\begin{table}
		\centering
%	\caption{Comparison of training time, test time and test accuracy on the multi-class classification tasks}
	% Table generated by Excel2LaTeX from sheet '统一参数全部算法'
		\begin{tabular}{l|cccccc}
	\hline
	% Table generated by Excel2LaTeX from sheet 'Sheet1'
	\multirow{2}[0]{*}{Algorithm} & \multicolumn{3}{c}{combined} & \multicolumn{3}{c}{mnist600k} \\
	& Train(s) & Test(s) & TestAcc(\%) & Train(s) & Test(s) & TestAcc(\%) \\
%	\begin{tabular}{|l|r|r|r|r|r|r|}
%		\hline
%		\multicolumn{1}{|c|}{\multirow{2}[4]{*}{Algorithm}} & \multicolumn{3}{c|}{combined} & \multicolumn{3}{c|}{mnist600k}\\
%		\cline{2-7}    & \multicolumn{1}{c|}{Train(s)} & \multicolumn{1}{c|}{Test(s)} & \multicolumn{1}{c|}{TestAcc} & \multicolumn{1}{c|}{Train(s)} & \multicolumn{1}{c|}{Test(s)} & \multicolumn{1}{c|}{TestAcc} \\
		\hline
		LR &      3.25  &      0.17  &    73.563  &     22.73  &      1.04  &    53.961  \\
		
		OGD &      2.85  &      0.17  &    77.792  &     19.53  &      1.04  &    83.743  \\
		
		PA &      3.49  &      0.17  &    75.936  &     23.00  &      1.04  &    82.630  \\
%		\hline
		FOGD &      8.60  &      1.52  &    79.249  &    101.60  &     20.88  &    95.680  \\
		
		NOGD &     38.68  &      3.98  &    71.526  &    194.88  &     22.40  &    78.585  \\
		
		RRF &     66.35  &      2.89  &    79.995  &    361.01  &     39.36  &    95.712  \\
		
		AVM &     10.59  &      2.04  &    60.601  &     55.85  &      9.92  &    84.399  \\
		
		MPU-FOGDU &    117.68  &      2.19  &    83.223  &    126.21  &     18.43  &    99.096  \\
		
		MPU-FOGDUB &    121.26  &      2.20  &    83.214  &    137.20  &     18.48  &    99.097  \\
		\hline
	\multirow{2}[0]{*}{Algorithm} & \multicolumn{3}{c}{poker} & \multicolumn{3}{c}{acoustic} \\
& Train(s) & Test(s) & TestAcc(\%) & Train(s) & Test(s) & TestAcc(\%) \\
%		\multicolumn{1}{|c|}{\multirow{2}[4]{*}{Algorithm}} & \multicolumn{3}{c|}{poker} & \multicolumn{3}{c|}{acoustic} \\
%		\cline{2-7}    & \multicolumn{1}{c|}{Train(s)} & \multicolumn{1}{c|}{Test(s)} & \multicolumn{1}{c|}{TestAcc} & \multicolumn{1}{c|}{Train(s)} & \multicolumn{1}{c|}{Test(s)} & \multicolumn{1}{c|}{TestAcc} \\
		\hline
		LR &     40.25  &      1.69  &    27.160  &      3.22  &      0.17  &    67.039  \\
		
		OGD &     38.41  &      1.68  &    31.614  &      3.09  &      0.17  &    67.335  \\
		
		PA &     45.80  &      1.68  &    28.331  &      3.66  &      0.17  &    65.489  \\
%		\hline
		FOGD &    207.22  &     37.45  &    48.089  &      9.04  &      1.55  &    67.658  \\
		
		NOGD &    281.95  &     26.91  &    52.290  &     31.47  &      3.91  &    67.500  \\
		
		RRF &    153.10  &      6.85  &    48.702  &     10.82  &      0.60  &    67.438  \\
		
		AVM &     70.21  &     11.22  &    44.396  &      8.92  &      1.60  &    58.972  \\
		
		MPU-FOGDU &    304.61  &     32.07  &    93.658  &     63.16  &      2.16  &    74.032  \\
		
		MPU-FOGDUB &    347.70  &     32.30  &    94.270  &     64.66  &      2.11  &    74.304  \\
		\hline
	\multirow{2}[0]{*}{Algorithm} & \multicolumn{3}{c}{forest} & \multicolumn{3}{c}{aloi} \\
& Train(s) & Test(s) & TestAcc(\%) & Train(s) & Test(s) & TestAcc(\%) \\
%		\multicolumn{1}{|c|}{\multirow{2}[4]{*}{Algorithm}} & \multicolumn{3}{c|}{forest} & \multicolumn{3}{c|}{aloi} \\
%		\cline{2-7}    & \multicolumn{1}{c|}{Train(s)} & \multicolumn{1}{c|}{Test(s)} & \multicolumn{1}{c|}{TestAcc} & \multicolumn{1}{c|}{Train(s)} & \multicolumn{1}{c|}{Test(s)} & \multicolumn{1}{c|}{TestAcc} \\
		\hline
		LR &     20.74  &      0.99  &    51.347  &    100.34  &      0.56  &    58.419  \\
		
		OGD &     19.41  &      1.00  &    71.210  &     65.01  &      0.51  &    75.879  \\
		
		PA &     22.78  &      0.99  &    69.334  &     83.80  &      0.54  &    75.674  \\
%		\hline
		FOGD &    110.52  &     22.19  &    75.685  &     82.65  &      3.30  &    86.694  \\
		
		NOGD &    192.75  &     23.73  &    67.913  &    129.73  &      5.38  &    52.180  \\
		
		RRF &    931.69  &     37.56  &    75.298  &    209.03  &      9.11  &    86.778  \\
		
		AVM &     55.65  &     10.02  &    63.478  &    134.60  &     22.06  &    27.301  \\
		
		MPU-FOGDU &    238.76  &     19.46  &    84.884  &    178.55  &      2.52  &    83.576  \\
		
		MPU-FOGDUB &    279.44  &     19.14  &    84.951  &    178.91  &      2.33  &    83.597  \\
		\hline
	\end{tabular}%
\caption{\label{tab:multi}Comparison of training time, test time and test accuracy on the multi-classification tasks.}
\end{table}

We compare the proposed MPU-FOGD algorithm with online gradient descent algorithm (OGD) \cite{zinkevich2003online}, online passive aggressive algorithm (PA) \cite{crammer2006online}, online logistic regression (LR) \cite{berger1996maximum},Fourier online gradient descent algorithm (FOGD) \cite{wang2013large}, Nyström online gradient descent algorithm (NOGD) \cite{lu2016large}, reparameterized random feature algorithm (RRF) \cite{nguyen2017large}, hybrid the passive-aggressive strategy and the max-out function algorithm (PAMO) \cite{jorge2018passive}, and approximation vector machine (AVM) \cite{le2017approximation}. Except that OGD, PA, and LR are linear algorithms, the others are nonlinear. PAMO is a binary classification algorithm, so we don't do multi-class experiments for it in this study.

\textbf{\subsection{Experimental Setups}}

For all the algorithms, we use k-folder cross-validation to tune the parameters. For LR and OGD, the step size range is selected from $\{10^{(-7)}, 10^{(-6)},\cdots,10^{(-1)}\}$, while for PA, it is chosen from $\{2^{(-8)}, 2^{(-7)}, \cdots, 2^5\}$. 
We vary the Gaussian kernel width in the range of
$\{2^{(-16)},2^{(-14)},2^{(-12)},2^{(-10)},\\2^{(-8)},2^{(-6)},2^{(-4)},2^{(-2)},2^0,2^2, 2^4\}$.
For MPU-FOGD algorithm, $\eta _1$ is set to $100$, $\eta _2$ is  $0.1$ and $\eta _3$ is chosen from $\{10^{(-6)},10^{(-4)},10^{(-2)},10^{(-1)}\}$.
The reason why $\eta _1$ can be set to $100$ is that compared with other algorithms,such as FOGD, its gradient is multiplied by one in $D$.
For other algorithms, in nonlinear situations, the step size of gradient descent $\eta _1$ is set to $0.001$.
The features of data samples are normalized in the interval $[-1,1]$. For the binary classification algorithms, $D$ in the FOGD, RRF, and MPU-FOGD algorithms are selected from $\{200,300, 400, 500, 1000, 2000, 4000, 6000, 8000\}$. The rank $k$ in NOGD is set to $0.2B$, and the budget size $B$ is chosen through a random search in range $\{200,400,600,800,1000\}$. For multi-class algorithms, except for AVM and NOGD algorithms whose maximum budget value is set to $500$, other budget parameters are set the same range: $\{200,300,400,500,1000,2000\}$.
We randomly divide all of the data sets into a training set, a test set and a validation set, of which the test set and the validation set account for $20\%$ respectively, the remaining $60\%$ for training. In order to better evaluate the performance of the classifier, we use the test accuracy, training time and test time as the performance indicators.
%, yet $\eta _3$ was chosen from $\{10^{(-8)}, 10^{(-6)},10^{(-4)},10^{(-2)},10^{(-1)}\}$.

\textbf{\subsection{Experimental Results and Analysis}}

%The experimental results are listed in Tables 2 and 3 for binary classification and multi-class classification on the twelve diverse data sets respectively. 
The experimental results on twelve binary classification and multi-class classification data sets are listed in Tables 2 and 3, respectively.

From Tables 2 and 3, we find that compared with nonlinear learning algorithms, the linear learning algorithms hold faster training and test time. However, their accuracy is not satisfactory.
For nonlinear learning algorithms, 
in term of test accuracy, in most cases, MPU-FOGD is higher than the other algorithms. 

%In general, compared with MPU-FOGD, the algorithms with higher test accuracy often have longer training time and test time.
%(2) In term of test time, in most cases, MPU-FOGD is faster than the other algorithms. Compared with MPU-FOGD, the algorithms with shorter test time often have lower test accuracy.

\textbf{\subsection{Wilcoxon Signed-ranks Test}} 

We conduct the Wilcoxon signed-ranks test \cite{demvsar2006statistical} to check whether the proposed method is significantly better than the other algorithms. The test compares the performances of two algorithms $a$ and $b$ on multiple datasets. To run the test, we rank the differences in performances of two algorithms for each dataset. The differences are ranked according to their absolute values. The smallest absolute value obtains the rank 1, the second smallest gets the rank 2. In case of equality, average ranks are assigned. The statistics of the Wilcoxon signed-ranks test is defined as \cite{demvsar2006statistical,yang2014robust}:  
\begin{equation}
\begin{aligned}
z(a,b)=\frac {T(a,b)-n(n+1)/4}{\sqrt {n(n+1)(2n+1)/24}},
\end{aligned}
\end{equation}
where $n$ is the number of the datasets, $T(a,b)=min\{R^+ (a,b),\\ R^- (a,b)\}$. $R^+ (a,b)$ is the sum of the ranks for the datasets on which algorithm $a$ outperforms algorithm $b$, and $R^- (a,b)$ means the sum of the ranks for the opposite. They are defined as follows:
\begin{equation}
\begin{aligned}
&R^+(a,b)=\sum _{d_i>0}rank(d_i)+\frac12\sum _{d_i=0}rank(d_i)\\
&R^-(a,b)=\sum _{d_i<0}rank(d_i)+\frac12\sum _{d_i=0}rank(d_i)
\end{aligned}
\end{equation}
where $d_i$ is the accuracy difference between algorithms $a$ and $b$ on the $i$-th experimental dataset, $rank(d_i)$ is the rank value of the absolute $d_i$.

%We fix $b$ as MPU-FOGDU or MPU-FOGDUB, and let $a$ as one of compared algorithms. From Tables 2-3, based on formulas (11)-(12), we can obtain that except that the statistical result of $z$(RRF, MPU-FOGDUB) is $-2.7849$, the value of other Wilcoxon signed-ranks are $-3.0202$ for the accuracy statistics of MPU-FOGDU and MPU-FOGDUB. 
%This shows that in term of classification accuracy, MPU-FOGDU and MPU-FOGDUB are significantly better than the other algorithms with the significance level $\alpha=0.05$. In terms of test time, we can obtain the same conclusion for NOGD, RRF, and AVM.
We fix $b$ as MPU-FOGDUB, and let $a$ as one of compared algorithms. From Tables 2-3, based on formulas (13)-(14), we can obtain that $z$(FOGD, MPU-FOGDUB)=$z$(RRF, MPU-FOGDUB)=-$-2.16$, $z$(NOGD, MPU-FOGDUB)=$z$(AVM, MPU-FOGDUB)=$-3.06$. This shows that in term of test accuracy, MPU-FOGDUB, similarly as MPU-FOGDU, is significantly better than the other algorithms with the significance level $\alpha=0.05$.
% In terms of test time, we can obtain the same conclusion for FOGD, NOGD, RRF, and AVM.

\textbf{\section{Conclusions and Feature Work}}

In this paper, a multi-parameter updating Fourier online descent gradient algorithms (MPU-FOGD) are proposed to deal with large scale nonlinear classification problems. Theoretical analysis is provided to guarantee that the proposed random feature mapping can give a lower bound. Experimental results on several benchmark data sets demonstrate that compared with the state-of-the-art Fourier feature mapping online learning algorithms, the proposed MPU-FOGD can have better test accuracy. 

For very large data sets, distributed learning provides a way to solve privacy-protected problems. At the same time, it also settles the problem that a single machine cannot handle or needs to spend a lot of time. In the future work, we are going to apply the work in this paper to design large-scale distributed classification algorithms.

\bibliographystyle{alpha}
\bibliography{sample}

\newcommand{\etalchar}[1]{$^{#1}$}
\begin{thebibliography}{FBdC{\'A}RJ{\etalchar{+}}14}

\bibitem[BDPDP96]{berger1996maximum}
Adam Berger, Stephen~A Della~Pietra, and Vincent~J Della~Pietra.
\newblock A maximum entropy approach to natural language processing.
\newblock {\em Computational linguistics}, 22(1):39--71, 1996.

\bibitem[CDK{\etalchar{+}}06]{crammer2006online}
Koby Crammer, Ofer Dekel, Joseph Keshet, Shai Shalev-Shwartz, and Yoram Singer.
\newblock Online passive-aggressive algorithms.
\newblock {\em Journal of Machine Learning Research}, 7(Mar):551--585, 2006.

\bibitem[CGLL10]{chang2010tree}
Fu~Chang, Chien-Yang Guo, Xiao-Rong Lin, and Chi-Jen Lu.
\newblock Tree decomposition for large-scale svm problems.
\newblock {\em The Journal of Machine Learning Research}, 11:2935--2972, 2010.

\bibitem[CHL08]{chang2008coordinate}
Kai-Wei Chang, Cho-Jui Hsieh, and Chih-Jen Lin.
\newblock Coordinate descent method for large-scale l2-loss linear support
  vector machines.
\newblock {\em Journal of Machine Learning Research}, 9(Jul):1369--1398, 2008.

\bibitem[CTJ09]{cheng2009efficient}
Haibin Cheng, Pang-Ning Tan, and Rong Jin.
\newblock Efficient algorithm for localized support vector machine.
\newblock {\em IEEE Transactions on Knowledge and Data Engineering},
  22(4):537--549, 2009.

\bibitem[CYYX16]{chang2016semantic}
Xiaojun Chang, Yao-Liang Yu, Yi~Yang, and Eric~P Xing.
\newblock Semantic pooling for complex event analysis in untrimmed videos.
\newblock {\em IEEE transactions on pattern analysis and machine intelligence},
  39(8):1617--1632, 2016.

\bibitem[Dem06]{demvsar2006statistical}
Janez Dem{\v{s}}ar.
\newblock Statistical comparisons of classifiers over multiple data sets.
\newblock {\em Journal of Machine learning research}, 7(Jan):1--30, 2006.

\bibitem[DNSV19]{dhamecha2019between}
Tejas~Indulal Dhamecha, Afzel Noore, Richa Singh, and Mayank Vatsa.
\newblock Between-subclass piece-wise linear solutions in large scale kernel
  svm learning.
\newblock {\em Pattern Recognition}, 95:173--190, 2019.

\bibitem[FBdC{\'A}RJ{\etalchar{+}}14]{frias2014online}
Isvani Fr{\'\i}as-Blanco, Jos{\'e} del Campo-{\'A}vila, Gonzalo Ramos-Jimenez,
  Rafael Morales-Bueno, Agust{\'\i}n Ortiz-D{\'\i}az, and Yail{\'e}
  Caballero-Mota.
\newblock Online and non-parametric drift detection methods based on
  hoeffding’s bounds.
\newblock {\em IEEE Transactions on Knowledge and Data Engineering},
  27(3):810--823, 2014.

\bibitem[FS01]{fine2001efficient}
Shai Fine and Katya Scheinberg.
\newblock Efficient svm training using low-rank kernel representations.
\newblock {\em Journal of Machine Learning Research}, 2(Dec):243--264, 2001.

\bibitem[GJP18]{guo2018effective}
Yiqing Guo, Xiuping Jia, and David Paull.
\newblock Effective sequential classifier training for svm-based multitemporal
  remote sensing image classification.
\newblock {\em IEEE Transactions on Image Processing}, 27(6):3036--3048, 2018.

\bibitem[GMCR04]{gama2004learning}
Joao Gama, Pedro Medas, Gladys Castillo, and Pedro Rodrigues.
\newblock Learning with drift detection.
\newblock In {\em Brazilian symposium on artificial intelligence}, pages
  286--295. Springer, 2004.

\bibitem[HCL{\etalchar{+}}08]{hsieh2008dual}
Cho-Jui Hsieh, Kai-Wei Chang, Chih-Jen Lin, S~Sathiya Keerthi, and
  Sellamanickam Sundararajan.
\newblock A dual coordinate descent method for large-scale linear svm.
\newblock In {\em Proceedings of the 25th international conference on Machine
  learning}, pages 408--415, 2008.

\bibitem[Hoe63]{hoeffding1963probability}
Wassily Hoeffding.
\newblock Probability inequalities for sums of bounded random variables.
\newblock 58:13--30, 1963.

\bibitem[HST]{hare38s}
S~Hare, A~Saffari, and PH~Torr.
\newblock S. 2016. struck: Structured output tracking with kernels.
\newblock {\em IEEE Transactions on Pattern Analysis \& Machine Intelligence},
  38(10).

\bibitem[HWZ14]{hoi2014libol}
Steven~CH Hoi, Jialei Wang, and Peilin Zhao.
\newblock Libol: A library for online learning algorithms.
\newblock {\em The Journal of Machine Learning Research}, 15(1):495--499, 2014.

\bibitem[Joa06]{joachims2006training}
Thorsten Joachims.
\newblock Training linear svms in linear time.
\newblock In {\em Proceedings of the 12th ACM SIGKDD international conference
  on Knowledge discovery and data mining}, pages 217--226, 2006.

\bibitem[JP18]{jorge2018passive}
Javier Jorge and Roberto Paredes.
\newblock Passive-aggressive online learning with nonlinear embeddings.
\newblock {\em Pattern Recognition}, 79:162--171, 2018.

\bibitem[KD05]{keerthi2005modified}
S~Sathiya Keerthi and Dennis DeCoste.
\newblock A modified finite newton method for fast solution of large scale
  linear svms.
\newblock {\em Journal of Machine Learning Research}, 6(Mar):341--361, 2005.

\bibitem[KLD{\etalchar{+}}19]{kang2019grayscale}
Bin Kang, Dong Liang, Wan Ding, Huiyu Zhou, and Wei-Ping Zhu.
\newblock Grayscale-thermal tracking via inverse sparse representation-based
  collaborative encoding.
\newblock {\em IEEE Transactions on Image Processing}, 29:3401--3415, 2019.

\bibitem[KSBM01]{keerthi2001improvements}
S.~Sathiya Keerthi, Shirish~Krishnaj Shevade, Chiranjib Bhattacharyya, and
  Karuturi Radha~Krishna Murthy.
\newblock Improvements to platt's smo algorithm for svm classifier design.
\newblock {\em Neural computation}, 13(3):637--649, 2001.

\bibitem[LHW{\etalchar{+}}16]{lu2016large}
Jing Lu, Steven~CH Hoi, Jialei Wang, Peilin Zhao, and Zhi-Yong Liu.
\newblock Large scale online kernel learning.
\newblock {\em The Journal of Machine Learning Research}, 17(1):1613--1655,
  2016.

\bibitem[LL00]{li2000relaxed}
Yi~Li and Philip~M Long.
\newblock The relaxed online maximum margin algorithm.
\newblock In {\em Advances in neural information processing systems}, pages
  498--504, 2000.

\bibitem[LM01]{lee2001rsvm}
Yuh-Jye Lee and Olvi~L Mangasarian.
\newblock Rsvm: Reduced support vector machines.
\newblock In {\em Proceedings of the 2001 SIAM International Conference on Data
  Mining}, pages 1--17. SIAM, 2001.

\bibitem[LNNP16]{le2016dual}
Trung Le, Tu~Nguyen, Vu~Nguyen, and Dinh Phung.
\newblock Dual space gradient descent for online learning.
\newblock In {\em Advances in Neural Information Processing Systems}, pages
  4583--4591, 2016.

\bibitem[LNNP17]{le2017approximation}
Trung Le, Tu~Dinh Nguyen, Vu~Nguyen, and Dinh Phung.
\newblock Approximation vector machines for large-scale online learning.
\newblock {\em The Journal of Machine Learning Research}, 18(1):3962--4016,
  2017.

\bibitem[LSV08]{le2008bundle}
Quoc~V Le, Alex~J Smola, and Svn Vishwanathan.
\newblock Bundle methods for machine learning.
\newblock In {\em Advances in neural information processing systems}, pages
  1377--1384, 2008.

\bibitem[LWZ14]{lin2014sample}
Ming Lin, Shifeng Weng, and Changshui Zhang.
\newblock On the sample complexity of random fourier features for online
  learning: How many random fourier features do we need?
\newblock {\em ACM Transactions on Knowledge Discovery from Data (TKDD)},
  8(3):1--19, 2014.

\bibitem[MM01]{mangasarian2001lagrangian}
Olvi~L Mangasarian and David~R Musicant.
\newblock Lagrangian support vector machines.
\newblock {\em Journal of Machine Learning Research}, 1(Mar):161--177, 2001.

\bibitem[NGB17]{nanni2017handcrafted}
Loris Nanni, Stefano Ghidoni, and Sheryl Brahnam.
\newblock Handcrafted vs. non-handcrafted features for computer vision
  classification.
\newblock {\em Pattern Recognition}, 71:158--172, 2017.

\bibitem[NLBP17]{nguyen2017large}
Tu~Dinh Nguyen, Trung Le, Hung Bui, and Dinh~Q Phung.
\newblock Large-scale online kernel learning with random feature
  reparameterization.
\newblock In {\em IJCAI}, pages 2543--2549, 2017.

\bibitem[PV16]{pesaranghader2016fast}
Ali Pesaranghader and Herna~L Viktor.
\newblock Fast hoeffding drift detection method for evolving data streams.
\newblock In {\em Joint European conference on machine learning and knowledge
  discovery in databases}, pages 96--111. Springer, 2016.

\bibitem[Ros58]{rosenblatt1958perceptron}
Frank Rosenblatt.
\newblock The perceptron: a probabilistic model for information storage and
  organization in the brain.
\newblock {\em Psychological review}, 65(6):386--408, 1958.

\bibitem[RR08]{rahimi2008random}
Ali Rahimi and Benjamin Recht.
\newblock Random features for large-scale kernel machines.
\newblock In {\em Advances in neural information processing systems}, pages
  1177--1184, 2008.

\bibitem[SS00]{smola2000sparse}
Alex~J Smola and Bernhard Sch{\"o}lkopf.
\newblock Sparse greedy matrix approximation for machine learning.
\newblock 2000.

\bibitem[SSSS07]{Shalev2007Pegasos}
Shai Shalev-Shwartz, Yoram Singer, and Nathan Srebro.
\newblock Pegasos: Primal estimated sub-gradient solver for svm.
\newblock In {\em Machine Learning, Proceedings of the Twenty-Fourth
  International Conference (ICML 2007), Corvallis, Oregon, USA, June 20-24,
  2007}, 2007.

\bibitem[TKC05]{tsang2005core}
Ivor~W Tsang, James~T Kwok, and Pak-Ming Cheung.
\newblock Core vector machines: Fast svm training on very large data sets.
\newblock {\em Journal of Machine Learning Research}, 6(Apr):363--392, 2005.

\bibitem[TVSL10]{teo2010bundle}
Choon~Hui Teo, SVN Vishwanathan, Alex Smola, and Quoc~V Le.
\newblock Bundle methods for regularized risk minimization.
\newblock {\em Journal of Machine Learning Research}, 11(1), 2010.

\bibitem[WSH{\etalchar{+}}18]{wen2018efficient}
Zeyi Wen, Jiashuai Shi, Bingsheng He, Jian Chen, and Yawen Chen.
\newblock Efficient multi-class probabilistic svms on gpus.
\newblock {\em IEEE Transactions on Knowledge and Data Engineering},
  31(9):1693--1706, 2018.

\bibitem[WZH{\etalchar{+}}13]{wang2013large}
Jialei Wang, Peilin ZHAO, Steven~CH HOI, Jinfeng Zhuang, and Zhi-yong Liu.
\newblock Large scale online kernel classification.
\newblock 2013.

\bibitem[XZL{\etalchar{+}}18]{xu2018modeling}
Jun Xu, Wei Zeng, Yanyan Lan, Jiafeng Guo, and Xueqi Cheng.
\newblock Modeling the parameter interactions in ranking svm with low-rank
  approximation.
\newblock {\em IEEE Transactions on Knowledge and Data Engineering},
  31(6):1181--1193, 2018.

\bibitem[YDL19]{ye2019hybrid}
Rui Ye, Qun Dai, and MeiLing Li.
\newblock A hybrid transfer learning algorithm incorporating trsvm with gasen.
\newblock {\em Pattern Recognition}, 92:192--202, 2019.

\bibitem[YSZ{\etalchar{+}}18]{yao2018extracting}
Yazhou Yao, Fumin Shen, Jian Zhang, Li~Liu, Zhenmin Tang, and Ling Shao.
\newblock Extracting privileged information for enhancing classifier learning.
\newblock {\em IEEE Transactions on Image Processing}, 28(1):436--450, 2018.

\bibitem[YTH14]{yang2014robust}
Xiaowei Yang, Liangjun Tan, and Lifang He.
\newblock A robust least squares support vector machine for regression and
  classification with noise.
\newblock {\em Neurocomputing}, 140:41--52, 2014.

\bibitem[ZBMM06]{zhang2006svm}
Hao Zhang, Alexander~C Berg, Michael Maire, and Jitendra Malik.
\newblock Svm-knn: Discriminative nearest neighbor classification for visual
  category recognition.
\newblock In {\em 2006 IEEE Computer Society Conference on Computer Vision and
  Pattern Recognition (CVPR'06)}, volume~2, pages 2126--2136. IEEE, 2006.

\bibitem[Zha04]{zhang2004solving}
Tong Zhang.
\newblock Solving large scale linear prediction problems using stochastic
  gradient descent algorithms.
\newblock In {\em Proceedings of the twenty-first international conference on
  Machine learning}, page 116, 2004.

\bibitem[Zin03]{zinkevich2003online}
Martin Zinkevich.
\newblock Online convex programming and generalized infinitesimal gradient
  ascent.
\newblock In {\em Proceedings of the 20th international conference on machine
  learning (icml-03)}, pages 928--936, 2003.

\bibitem[ZOW16]{zhao2016person}
Rui Zhao, Wanli Oyang, and Xiaogang Wang.
\newblock Person re-identification by saliency learning.
\newblock {\em IEEE transactions on pattern analysis and machine intelligence},
  39(2):356--370, 2016.

\bibitem[ZSW{\etalchar{+}}19]{zhang2019visual}
Lihe Zhang, Jiayu Sun, Tiantian Wang, Yifan Min, and Huchuan Lu.
\newblock Visual saliency detection via kernelized subspace ranking with active
  learning.
\newblock {\em IEEE Transactions on Image Processing}, 29:2258--2270, 2019.

\bibitem[ZWZ{\etalchar{+}}17]{zuo2017distance}
Wangmeng Zuo, Faqiang Wang, David Zhang, Liang Lin, Yuchi Huang, Deyu Meng, and
  Lei Zhang.
\newblock Distance metric learning via iterated support vector machines.
\newblock {\em IEEE Transactions on Image Processing}, 26(10):4937--4950, 2017.

\bibitem[ZZH{\etalchar{+}}16]{zhou2016one}
Zhaoze Zhou, Wei-Shi Zheng, Jian-Fang Hu, Yong Xu, and Jane You.
\newblock One-pass online learning: A local approach.
\newblock {\em Pattern Recognition}, 51:346--357, 2016.

\end{thebibliography}

\end{document}